\documentstyle[12pt,cite]{article}
\begin{document}\bibliographystyle{plain}\begin{titlepage}
\renewcommand{\thefootnote}{\fnsymbol{footnote}}\hfill\begin{tabular}{l}
HEPHY-PUB 775/03\\UWThPh-2003-21\\September 2003\end{tabular}\\[4cm]\Large
\begin{center}{\bf THE CHARMED STRANGE MESON SYSTEM}\\\vspace{2cm}\large{\bf
Wolfgang LUCHA\footnote[1]{\normalsize\ {\em E-mail address\/}:
wolfgang.lucha@oeaw.ac.at}}\\[.3cm]\normalsize Institut f\"ur
Hochenergiephysik,\\\"Osterreichische Akademie der
Wissenschaften,\\Nikolsdorfergasse 18, A-1050 Wien, Austria\\[1cm]\large{\bf
Franz F.~SCH\"OBERL\footnote[2]{\normalsize\ {\em E-mail address\/}:
franz.schoeberl@univie.ac.at}}\\[.3cm]\normalsize Institut f\"ur Theoretische
Physik, Universit\"at Wien,\\Boltzmanngasse 5, A-1090 Wien, Austria\vfill
{\normalsize\bf Abstract}\end{center}\normalsize

Motivated by the recent surprising discovery of two new meson states with
$c\bar s$ quark content but unexpectedly low masses and narrow total decay
widths, we work out,~in a nonrelativistic potential-model approach developed
already some two decades ago, the predictions for the energy levels of the
corresponding charm--anti-strange quark bound states. In spite of the fact
that this simple quark model reproduces the mass spectrum of the previously
observed hadrons remarkably well, we are led to the conclusion that, without
considerable modifications, both the new states do not fit into this
framework.\vspace{9ex}

\noindent{\em PACS numbers\/}: 14.40.Lb, 14.40.Ev, 12.39.Jh, 12.39.Pn
\renewcommand{\thefootnote}{\arabic{footnote}}\end{titlepage}

\normalsize

\section{Introduction}Recently, evidence for the existence of two new mesonic
states with charm and strange quark content has been presented by three
experiments \cite{BaBar03,CLEO03,CLEO03spcd,PompiliQCD03,Belle03,Belle03a,
RobuttiEPS03,VogelEPS03,SeusterEPS03,PorterEPS03,Belle03b,Stone03,Palano03}.
Table~\ref{Tab:Expts} summarizes~the experimental measurements of the masses
of the two new states as well as our weighted averages of the meson masses.
The uncertainties of the averages have been obtained by adding in quadrature
the statistical and systematic errors reported by the experiments.

At first, the {\sc BaBar} Collaboration \cite{BaBar03}, running the {\sc
BaBar} detector at the PEP-II asymmetric-energy ${\rm e}^+{\rm e}^-$ storage
ring, announced the observation of a charmed~strange meson called
$D_{sJ}^*(2317)^\pm.$ This new state manifested as narrow resonance with a
mass of $M(D_{sJ}^*(2317)^\pm)=(2316.8\pm0.4\pm3.0)\;\mbox{MeV}$ in the
$D_s^\pm\pi^0$ invariant-mass distribution.

The CLEO Collaboration \cite{CLEO03,CLEO03spcd,Stone03} provided the
confirmation of the existence of the $D_{sJ}^*(2317)^\pm$ meson from data
collected with the CLEO II detector in symmetric ${\rm e}^+{\rm e}^-$
collisions at the Cornell Electron Storage Ring. Furthermore, the CLEO
Collaboration \cite{CLEO03,CLEO03spcd,Stone03} reported the observation of a
charmed, strange meson called $D_{sJ}(2463)^\pm.$~This new state manifested
itself as narrow resonance in the $D_s^{*\pm}\pi^0$ invariant-mass spectrum.
Interestingly, the CLEO experiment \cite{CLEO03spcd,Stone03} found, within
errors, identical results for the mass differences (with statistical and
systematic errors) between these two $D_{sJ}^*(2317)^\pm$ and
$D_{sJ}(2463)^\pm$ states and the well-established \cite{PDG02} $D_s^\pm$ and
$D_s^{*\pm}$ mesons, respectively:\begin{eqnarray*}
M(D_{sJ}^*(2317)^\pm)-M(D_s^\pm)&=&(350.0\pm1.2\pm1.0)\;\mbox{MeV}\ ,\\[1ex]
M(D_{sJ}(2463)^\pm)-M(D_s^{*\pm})&=&(351.2\pm1.7\pm1.0)\;\mbox{MeV}\ .
\end{eqnarray*}With the most recent Particle Data Group averages \cite{PDG02}
$M(D_s^\pm)=(1968.1\pm0.5)\;\mbox{MeV},$ for the mass of the $D_s^\pm$ meson,
and $M(D_s^{*\pm})=(2111.9\pm0.7)\;\mbox{MeV},$ for the mass of~the
$D_s^{*\pm}$ meson, the mass differences translate into the resonance masses
listed in Table~\ref{Tab:Expts}. In return the {\sc BaBar} experiment
confirmed the existence of the $D_{sJ}(2463)^\pm$ meson \cite{RobuttiEPS03}.

Finally, both discoveries have been confirmed by the Belle Collaboration,
operating the Belle detector at the KEKB asymmetric-energy ${\rm e}^+{\rm
e}^-$ collider; the Belle experiment identified the new $D_{sJ}^*(2317)^\pm$
and $D_{sJ}(2463)^\pm$ states in the exclusive $B$ meson decays $B\to
D+D_{sJ}^*(2317)$ and $B\to D+D_{sJ}(2463)$ \cite{Belle03,Belle03b}, and
continuum ${\rm e}^+{\rm e}^-$ annihilations \cite{Belle03a}.

\small\begin{table}[hbt]\caption{Masses of the two new narrow charmed,
strange meson states $D_{sJ}^*(2317)^\pm$ and $D_{sJ}(2463)^\pm$ with both
statistical and systematic errors, measured by the experiments {\sc BaBar}
\cite{BaBar03,RobuttiEPS03}, CLEO
\cite{CLEO03,CLEO03spcd,VogelEPS03,Stone03}, and Belle
\cite{Belle03,Belle03a,SeusterEPS03,Belle03b}, as well as their weighted
averages}\label{Tab:Expts}
\begin{center}\begin{tabular}{llll}\hline\hline\\[-1.5ex]
\multicolumn{1}{c}{Experiment}&\multicolumn{1}{c}{Process}&
\multicolumn{1}{c}{$M(D_{sJ}^*(2317)^\pm)\ [\mbox{MeV}]$}&
\multicolumn{1}{c}{$M(D_{sJ}(2463)^\pm)\
[\mbox{MeV}]$}\\[1ex]\hline\\[-1.5ex]{\sc BaBar}
\cite{BaBar03,RobuttiEPS03}&inclusive ${\rm e}^+{\rm e}^-$&
$2316.8\pm0.4\pm3.0$&$2457.0\pm1.4\pm3.0$\\[1ex]CLEO
\cite{CLEO03spcd,VogelEPS03,Stone03}&inclusive ${\rm e}^+{\rm e}^-$&
$2318.1\pm1.2\pm1.0$&$2463.1\pm1.7\pm1.0$\\[1ex]Belle
\cite{Belle03,Belle03b}&$B$ decays&
$2319.8\pm2.1\pm2.0$&$2459.2\pm1.6\pm2.0$\\[1ex]Belle
\cite{Belle03a}&inclusive ${\rm e}^+{\rm e}^-$&
$2317.2\pm0.5\pm0.9$&$2456.5\pm1.3\pm1.1$\\[1ex]\hline\\[-1.5ex]
\multicolumn{2}{c}{Weighted Averages}&$2317.6\pm0.8$&$2459.0\pm1.1$
\\[1ex]\hline\hline\end{tabular}\end{center}\end{table}\normalsize

Until now, for the $D_{sJ}^*(2317)^\pm$ state, only the decay
$D_{sJ}^*(2317)^+\to D_s^++\pi^0$ and, for the $D_{sJ}(2463)^\pm$ meson, the
decays $D_{sJ}(2463)^+\to D_s^{*+}+\pi^0$ and $D_{sJ}(2463)^+\to
D_s^++\gamma$ (as well as, of course, their charge-conjugate channels) have
been observed \cite{BaBar03,CLEO03,CLEO03spcd,PompiliQCD03,Belle03,Belle03a,
RobuttiEPS03,VogelEPS03,SeusterEPS03,Belle03b,Stone03}. For both states, the
widths of the resonance peaks are consistent with the experimental mass
resolution. This indicates that the observed resonance widths are consistent
with those expected for states of zero intrinsic or natural width $\Gamma.$
The CLEO Collaboration reports identical upper limits:
$\Gamma(D_{sJ}^*(2317)^\pm)<7\;\mbox{MeV}$ for the $D_{sJ}^*(2317)^\pm$ meson
and $\Gamma(D_{sJ}(2463)^\pm)<7\;\mbox{MeV}$ for the $D_{sJ}(2463)^\pm$
meson, at the 90\% confidence level \cite{CLEO03spcd,Stone03}.

Concerning the quantum numbers of these two narrow resonances, all
experimental findings, such as all the observed decay modes and angular
distributions, are consistent with their interpretation as P-wave states with
spin-parity assignment $J^P=0^+$ for~the $D_{sJ}^*(2317)^\pm$ meson
\cite{BaBar03,CLEO03,CLEO03spcd,PompiliQCD03,Belle03a,RobuttiEPS03,VogelEPS03,
SeusterEPS03,Stone03} and $J^P=1^+$ for the $D_{sJ}(2463)^\pm$ meson
\cite{CLEO03,CLEO03spcd,PompiliQCD03,Belle03,Belle03a,RobuttiEPS03,VogelEPS03,
SeusterEPS03,Belle03b,Stone03,Palano03}.

The new states do not show the decay patterns favoured by theoretical
expectation:\begin{itemize}\item The mass of the $D_{sJ}^*(2317)^\pm$ meson
is $40.8\;\mbox{MeV}$ below the kinematical threshold $2358.4\;\mbox{MeV}$
for its isospin-conserving strong decay $D_{sJ}^*(2317)^+\to D^0+K^+,$ and
$49.4\;\mbox{MeV}$ below the kinematical threshold $2367.0\;\mbox{MeV}$ for
its isospin-conserving strong decay $D_{sJ}^*(2317)^+\to D^++K^0$
(charge-conjugate decays are understood).\item The mass of the
$D_{sJ}(2463)^\pm$ meson lies above the two kinematical thresholds~for the
decays $D_{sJ}(2463)^+\to D^0+K^+$ and $D_{sJ}(2463)^+\to D^++K^0;$ these
processes, however, are in conflict with the conservation of angular momentum
and parity.\item The mass of the $D_{sJ}(2463)^\pm$ meson is
$41.5\;\mbox{MeV}$ below the kinematical threshold $2500.5\;\mbox{MeV}$ for
its isospin-conserving decay $D_{sJ}(2463)^+\to D^*(2007)^0+K^+,$ and
$48.7\;\mbox{MeV}$ below the kinematical threshold $2507.7\;\mbox{MeV}$ for
its isospin-conserving decay $D_{sJ}(2463)^+\to D^*(2010)^++K^0$
(charge-conjugate decays are understood).\end{itemize}As a consequence of
their low masses blocking their isospin-conserving strong decays, the mesons
$D_{sJ}^*(2317)^\pm$ and $D_{sJ}(2463)^\pm$ decay via the channels
$D_{sJ}^*(2317)^+\to D_s^++\pi^0$ and $D_{sJ}(2463)^+\to D_s^{*+}+\pi^0,$
which both violate isospin conservation, and through~the radiative decay mode
$D_{sJ}(2463)^+\to D_s^++\gamma,$ as well as the respective charge-conjugate
modes. This explains the observed narrowness of the total decay widths of
these states.

The experimentally measured masses of the two $D_{sJ}^*(2317)^\pm$ and
$D_{sJ}(2463)^\pm$ states (summarized in Table~\ref{Tab:Expts}) are
significantly smaller than the corresponding predictions of most, but not
all, QCD-inspired quark potential models, which regard these mesons as P-wave
bound states $(c\bar s)$ of a charm quark $c$ and a strange antiquark $\bar
s.$ Concerning the masses of the $J^P=0^+$ and the lower $J^P=1^+$ $(c\bar
s)$ states, typical sets of predictions~are obtained, e.g., in the
comprehensive potential-model analyses of
Refs.~\cite{Godfrey85,Godfrey91,DiPierro01}, whereas results at the low-mass
end of the spectrum of predictions can be found in
Refs.~\cite{Gupta95,Zeng95}. The failure of quark potential models to explain
such low masses of the new resonances prompted a multitude of speculations
about the genuine nature of these states
\cite{Cahn03,Barnes03,Beveren03,Cheng03,Bardeen03,Szczepaniak03,Godfrey03,
Colangelo03,Bali03,Terasaki03,Rosner03,Beveren03a,Beveren03TJF,Nussinov03,
Lipkin03,Arkhipov03,Dai03,Rosner03a,Dougall03,Browder03,Deandrea03,Chen03,
Sadzikowski03,Nowak03,Datta03,Suzuki03,Kolomeitsev03,Cheng03a,Bali03a,
Hofmann03,Tornqvist03,Vijande03,Bianco03,Terasaki03a,Cohen03,Terasaki03b,
Fayyazuddin03,Vijande03a,Arkhipov03a}. Tentative interpretations view the
mesons $D_{sJ}^*(2317)^\pm$ and $D_{sJ}(2463)^\pm$ dominantly as $(c\bar s)$
bound states \cite{Cahn03,Beveren03,Bardeen03,Godfrey03,Colangelo03,Bali03,
Beveren03a,Dai03,Dougall03,Kolomeitsev03,Hofmann03,Vijande03}, $(DK)$
molecules \cite{Barnes03}, four-quark bound states
\cite{Cheng03,Terasaki03,Nussinov03,Terasaki03a}, or $D\pi$ atoms
\cite{Szczepaniak03}. Clearly, in general each of these mesons is a linear
combination of all Fock states which carry the appropriate quantum numbers.

In this work, we examine whether the new $D_s$ mesons can be accommodated
as~$(c\bar s)$ bound states in a phenomenologically appealing nonrelativistic
quark potential model. After recalling in Sec.~\ref{Sec:Spectre} the
spectroscopic classification of heavy-light states, we sketch in
Sec.~\ref{Sec:Model-N} the adopted quark model, and present in
Sec.~\ref{Sec:R&C} our---disappointing---findings.

\section{Spectroscopy of heavy-quark light-quark mesons}\label{Sec:Spectre}
The spectroscopic classification of bound states $(Q\bar q)$ composed of a
heavy quark $Q$ and a comparatively light antiquark $\bar q$ is greatly
facilitated by the following observation.~In the limit of infinitely large
mass of the heavy quark $Q,$ the spin $\mbox{\boldmath{$s$}}_Q$ of the
heavy~quark $Q$ and the total angular momentum $\mbox{\boldmath{$j$}}$ of the
light quark $q,$ defined as the sum
$\mbox{\boldmath{$j$}}=\mbox{\boldmath{$l$}}+\mbox{\boldmath{$s$}}_q$ of the
orbital angular momentum $\mbox{\boldmath{$l$}}$ and the spin
$\mbox{\boldmath{$s$}}_q$ of the light quark $q,$ are separately conserved.
Hence, in this case $\mbox{\boldmath{$j$}}$ provides a ``good'' quantum
number for classification.

The light-quark angular momentum $\mbox{\boldmath{$j$}}$ and the spin
$\mbox{\boldmath{$s$}}_Q$ of the heavy quark combine to the total angular
momentum
$\mbox{\boldmath{$J$}}=\mbox{\boldmath{$j$}}+\mbox{\boldmath{$s$}}_Q$ of the
$(Q\bar q)$ bound state. The corresponding quantum number $J$ fixes the spin
of the resulting meson. The parity $P$ of this meson is related to the
quantum number $\ell$ of the orbital angular momentum $\mbox{\boldmath{$l$}}$
by $P=(-1)^{\ell+1}.$

Thus, in the heavy-quark limit the heavy-light bound states $(Q\bar q)$ may
conveniently be classified in terms of the light-quark total-angular-momentum
quantum number $j$:\begin{description}\item[S-wave states] have orbital
angular momentum $\ell=0$ and, therefore, negative parity, $P=-1.$ Spin and
orbital angular momentum of the light quark $q$ can couple~only to
$j^P=\frac{1}{2}^-.$ Combining the light-quark total angular momentum
$\mbox{\boldmath{$j$}}=\mbox{\boldmath{$s$}}_q$ with the spin of the heavy
quark $Q$ yields a spin-singlet state with spin-parity assignment $J^P=0^-$
and a spin-triplet state with spin-parity assignment $J^P=1^-.$

The pseudoscalar $J^P=0^-$ state has been identified with the isosinglet
$D_s^\pm$~meson (the former $F^\pm$ meson), with a mass of
$(1968.1\pm0.5)\;\mbox{MeV}$ and a well-established spin-parity assignment
$J^P=0^-$ \cite{PDG02}. The vector $J^P=1^-$ state is assumed~to~be identical
to the isosinglet $D_s^{*\pm}$ meson, with a mass of
$(2111.9\pm0.7)\;\mbox{MeV},$ natural spin-parity, and width and decay modes
consistent with the assignment $J^P=1^-$ \cite{PDG02}.\item[P-wave states]
have orbital angular momentum $\ell=1$ and, therefore, positive parity,
$P=+1.$ Spin and orbital angular momentum of the light quark can
couple~either to $j^P=\frac{1}{2}^+$ or to $j^P=\frac{3}{2}^+.$ Combination
with the spin of the heavy quark~yields, for $j^P=\frac{1}{2}^+,$ two states
with spin-parity assignment $J^P=0^+$ and $J^P=1^+,$~and, for
$j^P=\frac{3}{2}^+,$ two states with spin-parity assignment $J^P=1^+$
and~$J^P=2^+.$~The two $J^P=1^+$ states do not have definite
charge-conjugation properties; therefore, they can undergo mixing.

The vector $J^P=1^+$ state belonging to the $j^P=\frac{3}{2}^+$ doublet (with
possibly~small admixtures of its vector $J^P=1^+$ counterpart belonging to
the $j^P=\frac{1}{2}^+$ doublet) is, in general, assumed to be identical to
the isosinglet $D_{s1}(2536)^\pm$ meson, with~a mass of
$(2535.35\pm0.34\pm0.5)\;\mbox{MeV}$ and its spin-parity assignment
$J^P=1^+$~which is strongly favoured but still needs confirmation
\cite{PDG02}.

The spin-triplet tensor $J^P=2^+$ state is identified with the isosinglet
$D_{sJ}(2573)^\pm$ meson, with a mass of $(2572.4\pm1.5)\;\mbox{MeV},$
natural spin-parity, and width and decay modes consistent with the assignment
$J^P=2^+$ \cite{PDG02}.\end{description}Therefore, the present
phenomenological status of the $D_s$ system can be summarized~in the
following way: The two S-wave (ground) states are experimentally well
established. Good candidates for two of the four P-wave states predicted by
theory for bound states of two spin-$\frac{1}{2}$ fermions have been
observed. The open question is: May the $D_{sJ}^*(2317)^\pm$ and
$D_{sJ}(2463)^\pm$ states be interpreted as one or the other of the two
missing $(c\bar s)$ states?

\section{Nonrelativistic quark--antiquark potential~model}\label{Sec:Model-N}
The present analysis of the charmed strange meson system is based on a
nonrelativistic quark model improved by the lowest-order relativistic
corrections. For recent reviews~of the phenomenological description of
hadrons as bound states of quarks within the more intuitive framework of
(both nonrelativistic and semirelativistic) potential models see, for
instance, Refs.~\cite{Lucha91,Lucha92}. Since we intend to describe bound
states composed of both~a heavy quark and a light quark, we have to make use
of a framework which is capable~to reproduce the properties not only of heavy
quarkonia but also of light mesons \cite{Ono79,Ono82,Schoberl84}.

The nonrelativistic interquark potential $V(r)$ characterizing the
phenomenological model employed in the present analysis is of a
``funnel-like'' Coulomb-plus-linear shape, improved by an intermediate
exponential damping \cite{Ono79}. This quark model provides~an excellent
description of both the observed \cite{PDG02} meson and baryon mass spectra
\cite{Ono82,Schoberl84}. Apart from the strong fine-structure constant
$\alpha_{\rm s}$ and the masses of the quarks, this model involves six
constant parameters, $a,b,c,d,V_0,R_1,$ entering in the potential $V(r)$:
\begin{eqnarray*}V(r)&=&-\frac{4}{3}\,\frac{\alpha_{\rm
s}}{r}-b\exp{\left(-\frac{r}{c}\right)}+d+V_0\qquad\mbox{for}\ r\le R_1\
,\\[1ex]V(r)&=&a\,r-b\exp{\left(-\frac{r}{c}\right)}+V_0\qquad\mbox{for}\
r\ge R_1\ .\end{eqnarray*}From the smoothness of $V(r)$ at $r=R_1,$ the
potential parameters $R_1$ and $d$ are related to the strong fine-structure
constant $\alpha_{\rm s}$ and the slope $a$ of the linear term
according~to$$R_1=\sqrt{\frac{4\,\alpha_{\rm s}}{3\,a}}\ ,\quad
d=4\sqrt{\frac{\alpha_{\rm s}\,a}{3}}\ .$$Starting from the Bethe--Salpeter
formalism for the description of bound states within relativistic quantum
field theory, the static potential $V(r)$ in the Schr\"odinger equation may
be derived from the interaction kernel entering in the Bethe--Salpeter
equation by Fourier transformation. However, for a confining potential, such
as the linear potential $V(r)=a\,r,$ the corresponding Bethe--Salpeter kernel
has to be a highly singular object. The necessary regularization of the
kernel introduces a constant in the static potential. In order to compensate
the (infrared) divergence of the interaction kernel, in the case of the
linear potential $V(r)=a\,r$ this (negative) constant $V_0$ has to involve
the slope $a$ of the linear potential, and the Euler--Mascheroni constant
$\gamma=0.577\,215\,664\,901\dots$ \cite{Gromes81}:
$$V_0=-2\sqrt{a}\exp\left[-(\gamma-\mbox{$\frac{1}{2}$})\right]
\simeq-2\sqrt{a}\ .$$The above relations reduce the number of parameters in
the potential to four: $\alpha_{\rm s},a,b,c.$ With respect to the Lorentz
structure of the quark--antiquark interaction, the potential $V(r)$ is split
into a Lorentz-scalar part $V_{\rm S}=a\,r$ and a Lorentz-vector part $V_{\rm
V}=V-V_{\rm S}.$

The generalized Breit--Fermi Hamiltonian contains all relativistic
corrections to~the (nonrelativistic) Schr\"odinger Hamiltonian up to and
including the second order in the inverse masses of the bound-state
constituents. This lowest-order relativistic correction discriminates between
different Lorentz structures of the quark--antiquark interaction. The
spin-dependent relativistic corrections consist, for a Lorentz-vector spin
structure, of a spin--orbit, a spin--spin, and a tensor term, and, for a
Lorentz-scalar spin structure, of only a spin--orbit term. The spin--spin
term, however, involves the Laplacian $\Delta V_{\rm V}(r)$ of the vector
potential $V_{\rm V}(r);$ for any Coulomb-like contribution to the vector
potential $V_{\rm V}(r),$ i.e., a term in $V_{\rm V}(r)$ proportional to
$1/r,$ this yields a highly singular expression:
$$\Delta\frac{1}{r}=-4\pi\,\delta^{(3)}(\mbox{\boldmath{$r$}})\ .$$Similarly,
for a Coulomb-like contribution $V(r)\propto1/r$ to the nonrelativistic
interaction potential~$V(r)$ the spin--orbit and tensor terms introduce
singularities of the form~$1/r^3.$ This makes the Breit--Fermi Hamiltonian an
operator which is unbounded from below. However, for heavy bound-state
constituents, all relativistic corrections may be treated as perturbations of
the well-defined nonrelativistic Schr\"odinger Hamiltonian operator.

The mass difference between the pseudoscalar $\pi$ and the vector $\rho$
meson, generated by the spin--spin interaction, is of the same order of
magnitude as the involved meson masses. This indicates that, for any
description of mesons containing light quarks, the inclusion of at least the
spin--spin term in the nonperturbative treatment is mandatory. Within the
nonrelativistic framework, a conceivable solution is the ad-hoc replacement
of the $\delta$ function by a suitable smearing function $f(r)$ which
converges weakly towards the $\delta$ function in the appropriate limit,
i.e., the use of a representation of the $\delta$ function. This procedure
should be independent of the particular choice of the representation of the
$\delta$ function. The model of Ref.~\cite{Ono82} replaces the $\delta$
function by the smearing function
$$f(r)=\frac{1}{4\pi\,r_0^2}\,\frac{\exp{(-r/r_0)}}{r}\ ,$$satisfying
$$\lim_{r_0\to0}f(r)=\delta^{(3)}(\mbox{\boldmath{$r$}})\ .$$The dependence
of the smearing-range parameter introduced here, $r_0,$ on the masses of the
bound-state constituents is encoded in a purely phenomenological relationship
\cite{Ono82}. The numerical values of the parameters of this model have been
determined in Ref.~\cite{Ono82} from a fit of the mesons known at that time,
yielding, for the constituent quark masses, $m_{\rm u}=m_{\rm
d}=336\;\mbox{MeV},$ $m_{\rm s}=575\;\mbox{MeV},$ $m_{\rm
c}=1845\;\mbox{MeV},$ $m_{\rm b}=5235\;\mbox{MeV},$ and, for~the couplings in
the potential, $\alpha_{\rm s}=0.31,$ $a=0.15\;\mbox{GeV}^2,$
$b=0.956\;\mbox{GeV},$ $c=2.05\;\mbox{GeV}^{-1}.$

For this quark interaction, the mass eigenvalues and eigenstates of the
Breit--Fermi Hamiltonian are computed to high accuracy with the help of the
numerical integration procedure for the solution of the Schr\"odinger
equation developed in Ref.~\cite{Lucha99M}. However, before applying the
above quark model to the main target of the present investigation, viz., the
system of charmed strange ($D_s$) mesons, we would like to confront this
model's predictions for the entire mass spectrum of the experimentally
well-established mesons with results obtained in some representative
different approaches. For this comparison, we choose the relativized quark
potential model of Godfrey, Isgur, and Kokoski \cite{Godfrey85,Godfrey91},
and a relativistic quark model for mesons developed over the last decade by
some Bonn group within the Lorentz-covariant framework of the Bethe--Salpeter
equation \cite{Resag94,Muenz94,Koll00,Ricken00,Merten01,Ricken03}.

In order to get some idea of the significance of our approach, that is, to
estimate~the accuracy of our predictions, and to compare them with the
findings of different models, we introduce, as a measure of quality, a
quantity $Q,$ defined as the average of the square of the relative error of
the theoretical result $M_i^{\rm(th)}$ for the mass of the composite~particle
$i=1,2,3,\dots$ with respect to the corresponding experimentally measured
value $M_i^{\rm(exp)}$:$$Q\equiv\frac{1}{N}\sum_{i=1}^N
\left(\frac{M_i^{\rm(th)}-M_i^{\rm(exp)}}{M_i^{\rm(exp)}}\right)^2\ .$$The
smaller the numerical $Q$ value, the higher is the accuracy of the adopted
approach. For the numerical comparison of the models, we take into account
their predictions for $N=24$ energy levels corresponding to the
well-established mesons $\pi,$ $\rho(770),$ $\phi(1020),$ $\pi(1300),$
$\rho(1450),$ $K,$ $K^\ast(892),$ $D,$ $D^\ast(2010),$ $D_s,$ $D_s^\ast,$
$B,$ $B^\ast,$ $B_s,$ $\eta_c(1{\rm S}),$ $J/\psi(1{\rm S}),$ $\eta_c(2{\rm
S}),$ $\psi(2{\rm S}),$ $\psi(4040),$ $\psi(4415),$ $\Upsilon(1{\rm S}),$
$\Upsilon(2{\rm S}),$ $\Upsilon(3{\rm S}),$ and $\Upsilon(4{\rm S}).$
Numerically we~find for the quality measure $Q$ in our approach $Q=0.019\%,$
for the Godfrey--Isgur--Kokoski model \cite{Godfrey85,Godfrey91} $Q=0.041\%,$
which is roughly twice as large as our $Q$ value, and from the results
presented in Refs.~\cite{Koll00,Merten01} for the Bethe--Salpeter treatment
$Q=0.28\%,$ which~is more than an order of magnitude larger than our $Q$
value. From these numbers, we~feel entitled to conclude that this quark
potential model is definitely competitive (at least).

\section{Results, Discussion, and Conclusions}\label{Sec:R&C}
Table~\ref{Tab:Thy} gives the predictions of the quark model sketched in
Sec.~\ref{Sec:Model-N} for the masses~of~the lowest-lying six $(c\bar s)$
bound states expected by the spectroscopic classification of
Sec.~\ref{Sec:Spectre}. Notice that we still adopt the model parameter values
determined in the fit of Ref.~\cite{Ono82}. The formalism of the Breit--Fermi
Hamiltonian is more closely related to $L$-$S$ coupling. Consequently, the
states are labelled here by the usual spectroscopic notation
$n\,{}^{2\,S+1}\ell_J,$ where $n$ denotes the principal quantum number, $S$
is the quantum number of the total spin
$\mbox{\boldmath{$S$}}=\mbox{\boldmath{$s$}}_Q+\mbox{\boldmath{$s$}}_q,$ and
S, P, D, \dots\ indicate orbital angular momentum $\ell=0,1,2,\dots.$ There
are two $J^P=1^+$ states: the spin-singlet ${}^1{\rm P}_1$ state and the
spin-triplet ${}^3{\rm P}_1$ state. Because of non-diagonal contributions of
the relativistic corrections in the Breit--Fermi Hamiltonian, the two
$J^P=1^+$ states do not constitute mass eigenstates. The physical states
associated to the lower (l) or the higher (h) mass eigenvalue, respectively,
of~the Breit--Fermi Hamiltonian are called $D_{s1}^{\rm (l)}$ and
$D_{s1}^{\rm (h)}.$ The latter states are obtained from the states ${}^1{\rm
P}_1$ and ${}^3{\rm P}_1$ by an orthogonal transformation involving some
mixing angle~$\theta$:\begin{eqnarray*}|D_{s1}^{\rm (l)}\rangle&=&
\cos\theta\,|{}^1{\rm P}_1\rangle-\sin\theta\,|{}^3{\rm P}_1\rangle\
,\\[1ex]|D_{s1}^{\rm (h)}\rangle&=&
\sin\theta\,|{}^1{\rm P}_1\rangle+\cos\theta\,|{}^3{\rm P}_1\rangle\
.\end{eqnarray*}In our quark model we find for this ${}^1{\rm P}_1$-${}^3{\rm
P}_1$ mixing angle the numerical value $\theta=44.7^\circ.$

\small\begin{table}[hbt]\caption{Energy levels (in units of MeV) of the
lowest S- and P-wave states, identified by the usual spectroscopic notation
$n\,{}^{2\,S+1}\ell_J,$ of the charmed, strange quark--antiquark ($D_s$
meson) system predicted by the nonrelativistic quark model developed in
Ref.~\cite{Ono82}; the two mass eigenstates of the two-state $J^P=1^+$ P-wave
subsystem corresponding to lower and higher mass, formed by linear
combinations of the spin-singlet ${}^1{\rm P}_1$ state and the spin-triplet
${}^3{\rm P}_1$ state, are denoted by $D_{s1}^{\rm (l)}$ and $D_{s1}^{\rm
(h)},$ respectively. For comparison, the experimental values and some results
of typical quark models \cite{Godfrey91,Gupta95,Zeng95,DiPierro01} are
listed~too. Since Refs.~\cite{Godfrey91} and \cite{Zeng95} round their
results to $10\;\mbox{MeV}$ zeros are added where necessary. The
$D_{s1}(2536)^\pm$ of mass $(2535.35\pm0.6)\;\mbox{MeV}$ is one of the
$J^P=1^+$ states $D_{s1}^{\rm (l)}$ and~$D_{s1}^{\rm (h)}.$}\label{Tab:Thy}
\begin{center}\begin{tabular}{llllllll}\hline\hline\\[-1.5ex]
\multicolumn{1}{c}{State}&
\multicolumn{1}{c}{Experiment}&
\multicolumn{1}{c}{This Work}&
\multicolumn{1}{c}{Ref.~\cite{Godfrey91}}&
\multicolumn{1}{c}{Ref.~\cite{Gupta95}}&
\multicolumn{1}{c}{Ref.~\cite{Zeng95}}&
\multicolumn{1}{c}{Ref.~\cite{DiPierro01}}&
\multicolumn{1}{c}{Ref.~\cite{Merten01}}
\\[1ex]\hline\\[-1.5ex]
$1\,{}^1{\rm S}_0$&$1968.1\pm0.5$&1963&1980&1968.8&1940&1965&1969\\[1ex]
$1\,{}^3{\rm S}_1$&$2111.9\pm0.7$&2099&2130&2110.5&2130&2113&2116\\[1ex]
\hline\\[-1.5ex]
$1\,{}^3{\rm P}_0$&&2446&2484&2387.8&2380&2487&2464\\[1ex]
$D_{s1}^{\rm (l)}$&&2515&2550&2521.2&2510&2535&2506\\[1ex]
$D_{s1}^{\rm (h)}$&&2527&2560&2536.5&2520&2605&2506\\[1ex]
$1\,{}^3{\rm P}_2$&$2572.4\pm1.5$&2561&2590&2573.1&2580&2581&2552\\[1ex]
\hline\hline\end{tabular}\end{center}\end{table}\normalsize

Moreover, Table~\ref{Tab:Thy} compares the predictions of our approach with,
on the one hand, the experimentally observed masses \cite{PDG02} of all
previously known $D_s$ mesons and, on the other hand, the results of the four
(nonrelativistic) potential models
\cite{Godfrey85,Godfrey91,DiPierro01,Gupta95,Zeng95} explicitly mentioned in
the Introduction and of the Bonn Bethe--Salpeter model
\cite{Resag94,Muenz94,Koll00,Ricken00,Merten01,Ricken03}. First of all, we
note that all predictions of the Lorentz-covariant Bethe--Salpeter framework
do not substantially differ from the potential-model ones; this indicates
that the problems with the new $D_s$ states are not just a consequence of the
nonrelativistic approximation. Our results show a satisfactory agreement with
experiment and are within the range~of predictions of the other models. We
are forced to conclude that the model of Ref.~\cite{Ono82}, too, cannot
explain the masses of the new $D_s$ resonances; this discrepancy
becomes~still more distinct when inspecting the {\em binding energy\/} of the
quark--antiquark bound states.

We may get a clue to the correct identification of the heavier of the two
new~states, the $D_{sJ}(2463)^\pm$ meson, by inspection of the center of
gravity (COG) of the $\ell=1$ states. For P-wave states, the mass
$M(\mbox{COG})_{\rm P}$ of the center of gravity is defined
according~to$$M(\mbox{COG})_{\rm P}\equiv\frac{1}{9}\,[M({}^3{\rm
P}_0)+3\,M({}^3{\rm P}_1)+5\,M({}^3{\rm P}_2)]\ .$$Within the framework of
(nonrelativistic) interaction-potential models for two-particle bound states,
a {\em perturbative\/} inclusion of the spin--spin interaction implies the
equality of the P-wave center-of-gravity mass $M(\mbox{COG})_{\rm P}$ and the
mass $M({}^1{\rm P}_1)$ of the ${}^1{\rm P}_1$ state:$$M(\mbox{COG})_{\rm
P}=M({}^1{\rm P}_1)\qquad\mbox{(perturbatively)}\ .$$In contrast to this, in
a {\em nonperturbative\/} treatment of the spin--spin interaction by,~e.g.,
some smearing of Dirac's $\delta$ function---which should come closer to
reality---the P-wave center-of-gravity mass $M(\mbox{COG})_{\rm P}$ will be
larger than the mass $M({}^1{\rm P}_1)$ of the ${}^1{\rm P}_1$
state:$$M(\mbox{COG})_{\rm P}>M({}^1{\rm
P}_1)\qquad\mbox{(nonperturbatively)}\ .$$Let's neglect the mixing of the two
$J^P=1^+$ states necessary to get the physical states.\begin{itemize}\item
Identifying the ${}^3{\rm P}_1$ state with the $D_{s1}(2536)^\pm$ of mass
$(2535.35\pm0.34\pm0.5)\;\mbox{MeV},$ and the ${}^1{\rm P}_1$ state with the
$D_{sJ}(2463)^\pm$ meson of mass $(2459.0\pm1.1)\;\mbox{MeV},$
entails$$M(\mbox{COG})_{\rm P}^{\rm(exp)}=2531.7\;\mbox{MeV}>M({}^1{\rm
P}_1)=M(D_{sJ}(2463)^\pm)=2459.0\;\mbox{MeV}\ ,$$in accordance with the above
general theoretical expectations for this inequality.\item Identifying the
${}^3{\rm P}_1$ state with the $D_{sJ}(2463)^\pm$ meson of mass
$(2459.0\pm1.1)\;\mbox{MeV},$ and the ${}^1{\rm P}_1$ state with the
$D_{s1}(2536)^\pm$ of mass $(2535.35\pm0.34\pm0.5)\;\mbox{MeV},$
entails$$M(\mbox{COG})_{\rm P}^{\rm(exp)}=2506.3\;\mbox{MeV}<M({}^1{\rm
P}_1)=M(D_{s1}(2536)^\pm)=2535.35\;\mbox{MeV}\ ,$$in contradiction to the
above general theoretical expectation for this inequality.\end{itemize}This
simple observation may be regarded as a hint that the new $D_{sJ}(2463)^\pm$
resonance is predominantly a ${}^1{\rm P}_1$ state.

\end{document}